\documentclass[%
 aip,
 amsmath,amssymb,
 reprint,%
]{revtex4-1}
\usepackage{graphicx}
\usepackage{dcolumn}
\usepackage{bm}

\usepackage[utf8]{inputenc}
\usepackage[T1]{fontenc}
\usepackage{mathptmx}

\usepackage{nicefrac}       
\usepackage{microtype}      
\usepackage{mathtools}
\usepackage{commath}

\usepackage{float}
\usepackage{varioref}
\usepackage[sans]{dsfont}
\usepackage{tikz}

\usepackage[makeroom]{cancel}
\usepackage{graphicx}

\newcommand{\Mathematica}{\textit{Mathematica\textsuperscript{\resizebox{!}{0.8ex}{\textregistered}}}}
\usepackage{xspace}
\def\8{\infty}
\def\oh{\frac{1}{2}}
\def\ot{\frac{1}{3}}
\def\oq{\frac{1}{4}}
\def\tt{\frac{2}{3}}

\def\ofi{\frac{1}{5}}
\def\twfi{\frac{2}{5}}
\def\thfi{\frac{3}{5}}
\def\sifi{\frac{6}{5}}

\def\d{\partial}

\def\dal{\partial_{\alpha}}
\def\dbe{\partial_{\beta}}

\def\undertext#1{\vtop{\hbox{#1}\kern 1pt \hrule}}
\def\ra{\rightarrow}

\def\beq{\begin{equation}}
\def\eeq{\end{equation}}
\def\bea{\begin{eqnarray} & &}
\def\eea{\end{eqnarray}}

\def\EXP#1{\exp\left(#1\right)}

\def \R{\mbox{Re}}

\def\NS{Navier-Stokes}

\def\val{v_{\alpha}}
\def\vbe{v_{\beta}}

\def\XXint#1#2#3{{\setbox0=\hbox{$#1{#2#3}{\int}$}
     \vcenter{\hbox{$#2#3$}}\kern-.5\wd0}}

\DeclareMathOperator{\erf}{erf}

\begin{document}


\title{Asymmetric Vortex Sheet}
\author{Alexander Migdal}
\email{sasha.migdal@gmail.com}
\affiliation{Department of Physics, New York University \\
  726 Broadway, New York, NY 10003}

\begin{abstract}
We present a steady analytical solution of the incompressible Navier-Stokes equation for arbitrary viscosity in an arbitrary dimension $d$ of space. It represents a $d-1$ dimensional vortex "sheet" with an asymmetric profile of vorticity as a function of the normal coordinate $z$. This profile is related to the Hermite polynomials $H_\mu(z)$ which are analytically continued to the negative fractional index $\mu = -\frac{d}{d-1}$.
In $d=2$ dimensions, the solution degenerates to a constant vorticity flow. In $ d \ge 3$ dimensions, the vorticity is confined to the thin layer around the hyperplane with Gaussian decay on one side of the hyperplane and the power decay on another side. One can adjust the common scale of velocity so that the dissipation will stay finite at vanishing viscosity. In this limit, the width $w$ of the viscous lawyer will shrink to zero as $\nu^{\frac{3}{5}}$ for arbitrary dimension $d>3$. In $d=3$ dimensions, this power law is also accompanied by powers of the logarithm.
\end{abstract}

\maketitle
\section{\label{sec:level1} Introduction}

First, let us define the notations. We are using the Einstein convention of summation over repeated indexes.
The Greek indexes $\alpha,\beta,\dots $ with run from $1$ to $3$ and correspond to physical space $R_3$ and the lower case Latin indexes $a,b,\dots $ will take values $1,2$ and correspond to the internal parameters on the surface. 

We shall also use the Kronecker delta $\delta_{\alpha\beta}, \delta_{a b}$ in three and two dimensions as well as the antisymmetric tensors $e_{\alpha\beta\gamma}, e_{a b}$ normalized to $e_{1 2 3} = e_{1 2} =1$.

Let us define here the \NS{} equations , vorticity and the Hamiltonian.

\begin{subequations}
\begin{eqnarray}\label{NSv}
   && \d_t \vec v = \nu \vec \nabla^2\vec v - (\vec v \cdot \vec \nabla) \vec v - \vec \nabla p ;\\
   && \vec \nabla \cdot \vec v =0;\\
   &&\vec \omega = \vec \nabla \times \vec v ;\\
   && H = \int d^3 r \oh \vec v^2
\end{eqnarray}
\end{subequations}
The general identity, which follows from the steady \NS{} equation if one multiplies both sides by $\vec v$ and integrates over some volume $V$ in space:
\begin{align}
     & -\nu \int_V d^3 r \vec \omega^2 \nonumber \\
    &= \int_V d^3 r \dbe \left(\vbe \left(p + \oh \val^2\right) + \nu \val ( \dbe \val - \dal \vbe)\right)
\end{align}
By the Stokes theorem, the right side reduces to the flow over the boundary $\d V$ of the integration region $V$. The left side is the dissipation in this volume, so we find:
\begin{align}\label{enbalance}
    &\mathcal E =\nu \int_V d^3 r \vec \omega^2 = \nonumber\\
    &-\int_{\d V} d \vec \sigma \cdot \left(\vec v  \left(p + \oh \val^2\right) +\nu \vec \omega \times \vec v \right)
\end{align}

This identity holds for an arbitrary volume. 
The left side represents the viscous dissipation inside $V$, while the right side represents the energy flow through the boundary $\d V$.

This formula is equivalent to the conventional representation of dissipation through the trace of the square of the strain tensor:
\begin{eqnarray}
    S_{\alpha\beta} = \oh(\dal \vbe + \dbe \val)
\end{eqnarray}
The difference between the trace of the square of the vorticity vector and trace of the square of the strain tensor reduces to the total derivative terms, which we prefer to include in the definition of the energy flow through the boundary by means of Stokes theorem.

The Euler equation corresponds to setting $\nu =0$ in \eqref{NSv}. This limit is known not to be smooth, as the viscosity enters the term with the highest derivative.

Our goal here is to present an explicit steady solution with arbitrary viscosity, so that we can see what happens in the turbulent limit of vanishing viscosity at fixed energy dissipation.

This solution is a generalization of the Burgers vortex sheet \cite{BM05}, with the Gaussian profile of vorticity in the normal direction and a constant tangent discontinuity $\Delta v_y = 2 b$ in the Euler limit.
\begin{subequations}\label{BurgeSht}
\begin{eqnarray}
  &&\vec v = \{ - a x, b S_h(z), a z\};\\
  &&\vec \omega =\{-b S_h'(z),0,0\} ;\\
  &&S_h'(z) =\frac{1}{h \sqrt{2 \pi} } \EXP{- \frac{z^2}{2 h^2}};\\
  && S_h(z) =\erf \left( \frac{z}{h \sqrt 2}\right);\\
  && a = -\frac{\nu}{h^2};\label{aheq}
\end{eqnarray}
\end{subequations}

Our solution is different, being asymmetric and growing at infinity in one direction. It remains to be observed in real fluids or in DNS.

\section{Steady solution for the planar surface}

Let us consider an infinite plane at $z=0$ splitting the space $R_3$ into upper and lower half spaces.

We assume the separation of variables and write the following ansatz for the solution of the \NS{} equations
\begin{eqnarray}
    && v_i = S(z)\d_i \Gamma(x,y) + \d_i \Phi(x,y); \; i = 1,2\\
    && v_z = V(z);
\end{eqnarray}

We have from the incompressibility
\begin{eqnarray}
    V'(z)+ S(z) \d_i^2 \Gamma  + \d_i^2 \Phi=0;
\end{eqnarray}
This equation has the solution with linear $V(z)$, linear $\Gamma(x,y)$ and quadratic $\Phi(x,y)$. 
\begin{eqnarray}
    &&V(z) = a z;\\
    &&\Phi = -\oq a x_i^2;\\
    &&\Gamma(x,y) = q_i x_i
\end{eqnarray}
We wrote a \Mathematica{} code \cite{MB} to check the \NS{} equation in an arbitrary dimension of space.
In three dimensions, the \NS{} equation leads to the following hypergeometric equation for $S(z)$
\begin{equation}
      \nu S''  -  a z S' + \oh a S =0
\end{equation}

The velocity and pressure are now expressed  in terms of the solution $S(z)$ of this equation
\begin{eqnarray}
    && v_z = a z;\\\label{vz}
    && v_i = q_i S - \oh a x_i ;\\\label{vi}
    && p = -\frac{a^2 z^2}{2} - \frac{a^2x_i^2}{8},
\end{eqnarray}

For the negative $a$ there is a solution of this equation,  decaying in the upper half space (we used here \Mathematica{} to solve the equation and analyze the solution) :
\begin{eqnarray}
    &&a = -\frac{\nu}{w^2};\\
    &&S(z) = F_3\left(\frac{z}{w}\right);\\
    && F_d(z) = -\frac{2^{\frac{1}{2 (d-1)}} e^{-\frac{z^2}{2}} \Gamma \left(\frac{1}{d-1}\right) H_{-\frac{d}{d-1}}\left(\frac{z}{\sqrt{2}}\right)}{\sqrt{\pi }}\label{Fd}
\end{eqnarray}
where $ H_{\mu}(z)$ is the Hermite polynomial, analytically continued to the negative index $\mu$ using hypergeometric functions
\begin{eqnarray}\label{Hermite}
   &&H_\mu(z) = 2^{\mu} \sqrt{\pi}\frac{1}{\Gamma((1 - \mu)/2)}
{_1F_1}(-\mu/2, 1/2, z^2)\nonumber\\
&&-2^{\mu} \sqrt{\pi}\frac{2z}{\Gamma(-\mu/2)}{_1F_1}((1 - \mu)/2, 3/2, z^2))
\end{eqnarray}
The normalization of $S(z)$ is absorbed in the 2D vector parameter $\vec q$.
The parameter $w$ is arbitrary here, as well as $\vec q$.

\begin{figure}
    \centering
    \includegraphics[width=0.4\textwidth]{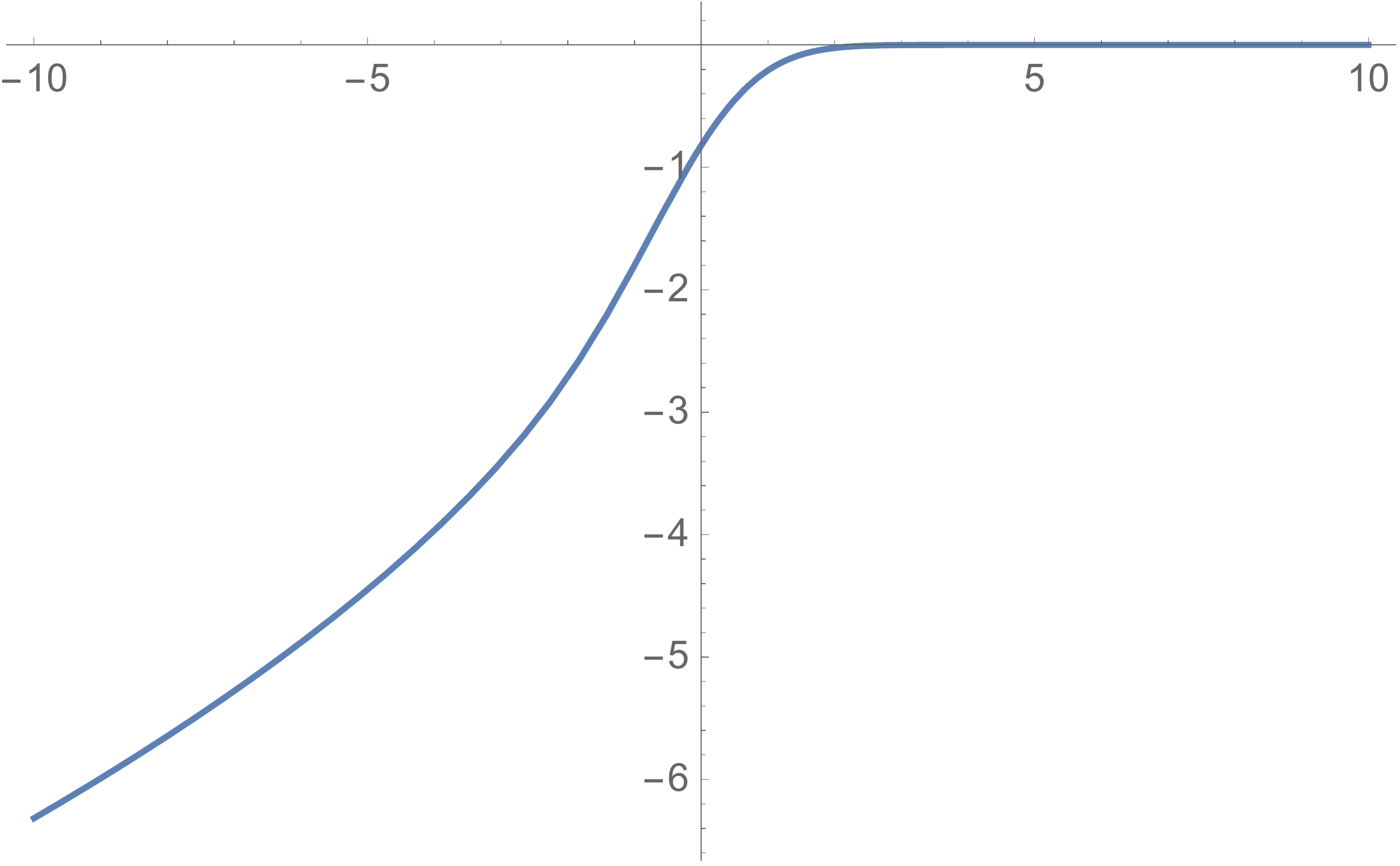}
    \caption{The velocity profile in normal direction.}
    \label{fig::Splot}
\end{figure}

Here is the plot of this function (Fig.\ref{fig::Splot}).

The resulting formula for vorticity $\vec \omega(x,y,z)$
\begin{eqnarray}
    &&\omega_z =0;\\
    &&\omega_x = -\frac{q_y}{w} F_3'\left(\frac{z}{w}\right);\\
    &&\omega_y = \frac{q_x}{w} F_3'\left(\frac{z}{w}\right);\\
\end{eqnarray}

At $z \ra +\infty$ , $F_3'(z)$ is exponentially small:
\begin{eqnarray}
    F_3'(z\ra  \infty) \ra  \frac{\EXP{-\frac{z^2}{2}}}{\sqrt{2z}}.
\end{eqnarray}

For negative $z$ it decays as a square root, up to exponential correction
\begin{eqnarray}
   F_3'(z\ra  -\infty) \ra \frac{1}{\sqrt{-z}}
\end{eqnarray}

\begin{figure}
    \centering
    \includegraphics[width=0.4\textwidth]{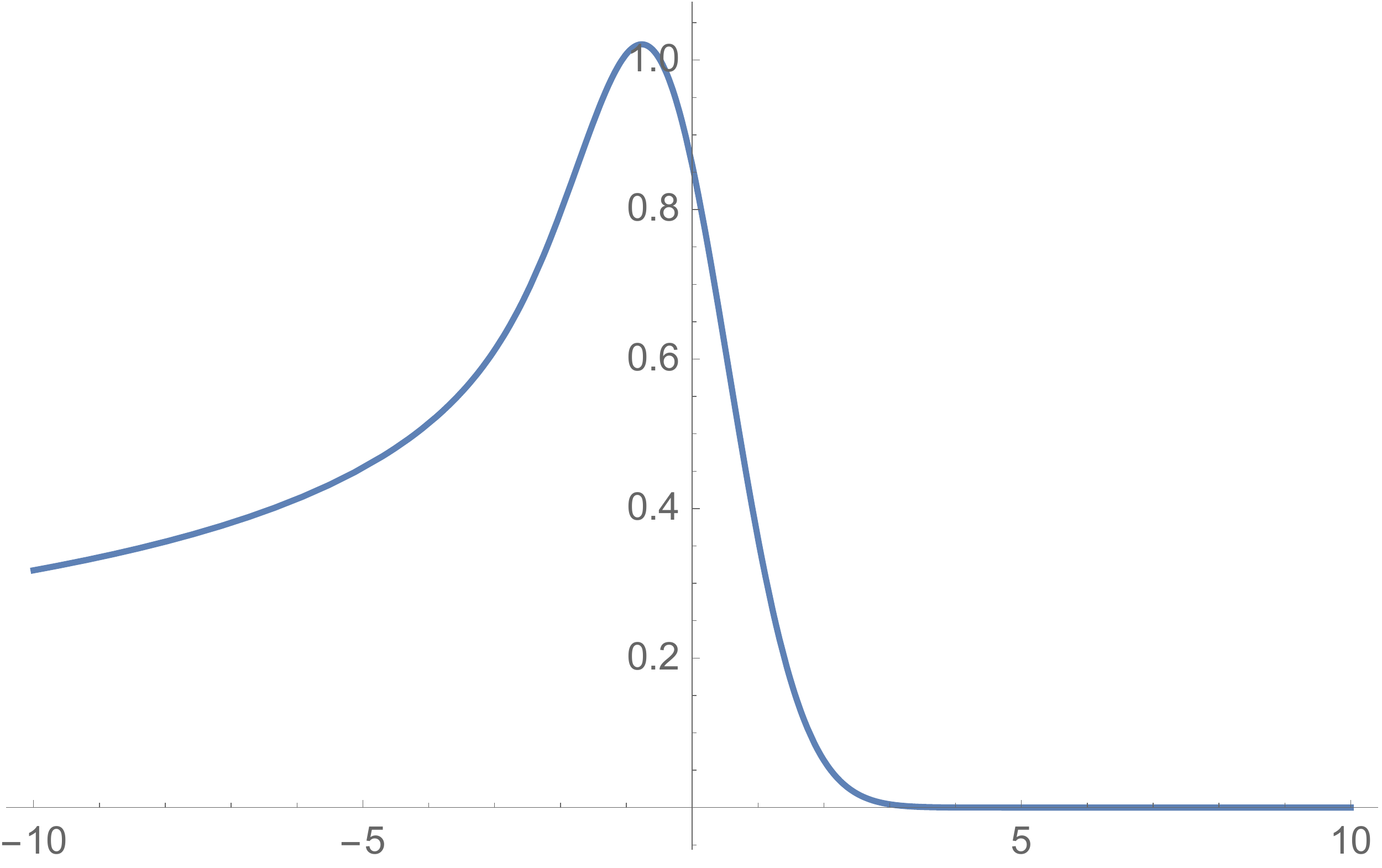}
    \caption{The vorticity profile in normal direction.}
    \label{fig::Wplot}
\end{figure}

Here is the plot of $F_3'(z)$ (Fig.\ref{fig::Wplot}).
It is asymmetric, with the maximum at $z = -0.764951$.

\section{Dissipation and the Turbulent limit}

The dissipation integral logarithmically diverges
\begin{eqnarray}
   &&\mathcal E = \nu \int_V d^3 r \vec \omega^2 = \nonumber\\
   &&\frac{\nu A q^2}{w} \left( \log (L/w)+ 1.6749 + O(1/L)\right) 
\end{eqnarray}
where $A$ is the area of the plane and $L$ is the depth of the lower half space.

We observe that the dissipation per unit area will be finite provided
\begin{eqnarray}\label{nurel}
   \frac{w}{\log (L/w)} \propto \nu \vec q^2;
\end{eqnarray}
On the other hand, we have another relation
\begin{equation}\label{adef}
    a = -\frac{\nu}{w^2}
\end{equation}

These two relations describe different parameters of the transverse velocity field inside the vortex sheet $z \sim w$
\begin{eqnarray}
   v_i \ra q_i F_3\left(\frac{z}{w}\right) - \oh a x_i
\end{eqnarray}
For both terms in the velocity to scale the same way in the vortex sheet at vanishing viscosity, we need to have
\begin{eqnarray}
   |\vec q| \propto |a| = \frac{\nu}{w^2}
\end{eqnarray}
Finally, plugging this formula into \eqref{nurel}, we find the following scaling relations between various parameters:
\begin{eqnarray}
   && \frac{w}{\log (L/w)} \propto \frac{\nu^3}{w^4} ;\\
   && \nu \propto w^{\frac{5}{3}}  \left(\log (L/w)\right)^{-\ot};\\
   && w \propto \nu^{\thfi} \left(\log (L/w)\right)^{\ofi};\\
   && |\vec v| \propto |a| \propto |\vec q| \propto \nu w^{-2} \propto w^{-\ot }\left(\log (L/w)\right)^{-\ot};\\
   && |\vec \omega| \propto  |\vec q| w^{-1} \propto w^{-\frac{4}{3} }\left(\log (L/w)\right)^{-\ot};\\
   && \R \propto \frac{a}{\nu} \propto w^{-2} \propto \nu^{-\sifi} \left(\log (L/w)\right)^{-\twfi}
\end{eqnarray}

These scaling laws are different from the ones suggested in my previous work\cite{M20c}, where I was assuming an ordinary vortex sheet with Gaussian profile.
The difference is only in the logarithmic factors, the scaling exponents are the same.

The outcome in the turbulent limit is  qualitatively the same: the velocity scale grows, the width of the vortex sheet shrinks, and the Reynolds number $\R$ grows in the turbulent limit. 

There is, however, one striking difference. There is no need for explicit random force. The energy pumped equals to the energy dissipated in any steady solution of the \NS{} equations, including ours. 

Explicit formula \eqref{enbalance} for the energy balance in this steady solution shows that all dissipated energy is pumped in from the boundaries (the walls of the large cube where we are considering our solution). Periodic boundary conditions are clearly impossible here, as there is a significant difference between fields in the upper and lower half-spaces.

\section{Solution for an arbitrary dimension}

This solution directly generalizes to an arbitrary space dimension $d$, with $z$ being the normal direction to the $d-1$ dimensional hyper-plane $x_i, i=1\dots d-1$.

The solution of steady \NS{} equations in $d$ dimensions reads \cite{MB}
\begin{eqnarray}
    && v_z = a z;\\\label{vzd}
    && v_i = q_i S - \frac{a x_i}{d-1};\\\label{vid}
    && p =-\frac{a^2 z^2}{2} -\frac{a^2 x_i^2}{2 (d-1)^2}
\end{eqnarray}
with $S(z)$ satisfying an equation
\begin{eqnarray}
    \nu S''  -  a z S' + \frac{a}{d-1}  S =0
\end{eqnarray}

With the same notation \eqref{adef} , \eqref{Fd},  we have now
\begin{eqnarray}
    S(z) = F_d\left(\frac{z}{w}\right);\\
\end{eqnarray}

It is worth noting that the Fourier transform $\tilde F_d(k) $ of this function $F_d(z)$ is an elementary function,satisfying the equation
\begin{eqnarray}
 &&\tilde F_d'(k) + \left(k  + \frac{d}{(d-1) k} \right) \tilde F_d(k) =0;\\
 && \tilde F_d(k)  = C k^{-\frac{d}{d-1}}\EXP{ - \frac{k^2}{2}};
\end{eqnarray}

Our solution with exponential decay at positive $z$ corresponds to a particular choice of the branch of the multivalued function $k^{-\frac{d}{d-1}}$, leading to the Hermite  polynomial with negative index \eqref{Hermite}.

Note that the Fourier components of velocity and vorticity are heavily peaked around zero with effective width $\sim 1/w$. This Gaussian tail is widening in the turbulent limit, being replaced by a power law.  

The singularity of the Fourier transform at $k =0$ is not integrable for any $ d>1$, which reflects the growth of velocity at large coordinates.  For the same reason, the total energy of this flow in a large box grows faster than volume. 

This is not a real obstacle, although, as the K41 scaling law also leads to growing velocity and the energy growing faster than volume. The difference is that here we have a velocity index $\oh$ instead of the $\ot$ for K41.

In two dimensions, this solution degenerates to $F_2(z) =z$. In this case, there is a constant vorticity and linear velocity in the whole plane. This is a good reminder that 2D turbulence is very different from the one we have in 3D.

At $d > 2$ this function decreases at large positive $z$  as $\EXP{-\frac{z^2}{2}}$. 
At large negative $z$ it grows as
\begin{equation}\label{Fdas}
F_d(z\ra -\infty) \ra -(d-1)(-z)^{\frac{1}{d-1}}
\end{equation}
 
 The corresponding antisymmetric vorticity tensor
 \begin{eqnarray}
 \omega_{\alpha\beta} = \dal\vbe-\dbe\val;\\
 \end{eqnarray}
 is vanishing unless when one of its two indexes equals $d-1$.
\begin{eqnarray}
 \omega_{(d-1) i} = -\omega_{i (d-1)} =  q_i S'(z) = \frac{q_i}{w} F'_d\left(\frac{z}{w}\right).
 \end{eqnarray}
 It does not depend upon the coordinates $x_i$ on the hyperplane, just as in the 3D case.

 Differentiating the \eqref{Fdas} for $S(z)$ we find for vorticity
 \begin{eqnarray}
  \omega_{(d-1) i}(z\ra -\infty) \ra \frac{q_i}{w} (-z)^{\frac{2-d}{d-1}}
 \end{eqnarray}
 
 This is a generalization of the square root decay we have in 3D.  Remarkably, $d=3$ is a special dimension in this theory. For $d>3$ the enstrophy integral converges at large negative $z$  
 \begin{eqnarray}
  \mathcal E =  \frac{\nu V_{d-1} q_i^2}{w}\int_{-\infty}^{\infty} d \xi (F'_d(\xi))^2 = \textit{const } \frac{\nu V_{d-1} q_i^2}{w}.
 \end{eqnarray}
 where $V_{d-1}$ is the volume of the hyperplane.
 
 As a consequence, at $d >3$ all scaling laws will stay without extra logarithmic factors.
 
 It is interesting that our space dimension is not only critical in quantum field theory, but also in the classical fluid dynamics.
 
 We live in a very special world, indeed.
 
\section{Conclusion}

We have found a  steady analytic solution of the \NS{} equation in an arbitrary dimension $d$ of space.
It represents a $d-1$ dimensional vortex "sheet" with an asymmetric profile of vorticity. On the upper side of the hyperplane it has an Gaussian decay, while on the lower side it decays as $|z|^{\frac{2-d}{d-1}}$.

In two dimensions, the solution degenerates to a constant vorticity.

In higher dimensions, the vorticity is confined to the narrow layer, with the width shrinking as a power of viscosity in the turbulent limit.
The $d=3$ is a special case, when the enstrophy integral logarithmically diverges on a lower side at large negative $z$.

The most remarkable thing is that it is an \textbf{exact} steady solution of the \NS{} equation for arbitrary viscosity and arbitrary space dimension. One could go from large viscosity to the turbulent limit of zero viscosity for a fixed energy flow and study the evolution of the flow along the way.

At this time, such asymmetric vortex sheets have not been observed yet, neither in real fluids nor in DNS. The ones observed are better explained by the symmetric Burgers solution.

\section*{Acknowledgments}

I am grateful to  Dmytro Bandak for useful discussions and comments.

This work is supported by a Simons Foundation award ID $686282$ at NYU. 

\bibliography{bibliography}

\end{document}